\documentclass[twocolumn,showpacs,preprintnumbers,amsmath,amssymb]{revtex4}

\usepackage{graphicx}
\usepackage{dcolumn}
\usepackage{bm}

\RequirePackage{cmap}
\RequirePackage[cp1251]{inputenc}
\RequirePackage[TS1,T2A]{fontenc}

\begin{document}

\title{Hot and dense gas of quark quasi-particles }

\author{S. V. Molodtsov}
 \altaffiliation[Also at ]{
Institute of Theoretical and Experimental Physics, Moscow, RUSSIA}
\affiliation{%
Joint Institute for Nuclear Research, Dubna,
Moscow region, RUSSIA
}%
\author{G. M. Zinovjev}
\affiliation{
Bogolyubov Institute for Theoretical Physics,
National Academy of Sciences of Ukraine, Kiev, UKRAINE
}%

\date{\today}

\begin{abstract}
 Some features of hot and dense gas of quarks which are considered as the
quasi-particles of the model Hamiltonian with four-fermion interaction are
studied. Being adapted to the Nambu-Jona-Lasinio model this approach allows us
to
accommodate a phase transition similar to the nuclear liquid-gas one at the
proper scale.
It allows us to argue an existence of the mixed phase of vacuum and baryonic
matter
(even at zero temperature) as a plausible scenario of chiral symmetry (partial)
restoration.

\end{abstract}

\maketitle

Understanding in full and describing dependably the critical phenomena (chiral
and deconfinement phase transitions) in QCD is still elusive because of a
necessity
to have the corresponding efficient non-perturbative methods for strongly
coupled regime to be analyzed. For the time being such studies are pursued by
invoking
diverse effective models. The Nambu-Jona-Lazinio(NJL)-type models are certainly
playing the most advanced role in this analysis \cite{Bub}. This approach deals
with the four-fermion interactions in lieu of a gluon field QCD dynamics and
does not incorporate (albeit being supplemented by the Polyakov loops it does
\cite{Fuku}) the property of confinement. In the meantime it is quite successful
in realizing the spontaneous breakdown of chiral symmetry and its restoration
at non-zero temperatures or quark densities. Apparently, the non-renormalizable
nature of the NJL model supposes another approximations introduced to be solved
and might lead to some conclusions which are dependent on the regularization
scenario sometimes. Hence, it requires a steadfast control of all inputs
done to have the consistent physical results and to avoid the reasons
for skeptical attitude.

Instructive example was given in Refs. \cite{MZ} and \cite{ff} in which the
ground state of model Hamiltonian with four-fermion interactions was studied in
detail.
The quarks were treated as the quasi-particles of this Hamiltonian and
unexpected
singularity (discontinuity) of the mean energy functional as a function of the
current quark
mass was found out. In particular case of the NJL model new solution branches of
the
equation for dynamical quark mass as a function of chemical potential (the
details are
shown below in Fig. \ref{f1}) have been found out, and the appearance of state
filled up
with quarks which is almost degenerate with the vacuum state both in quasi-
particle chemical
potential and in ensemble pressure has been discovered.

Here we are going to study the quark ensemble features at finite temperature and
fixed baryonic chemical potential and to analyse the first order phase
transition which takes place in such a system of free quasi-particles. Analysis
is
performed within the framework of two approaches which are supplementary, in a
sense,
albeit give the identical results. One of those approaches, based on the
Bogolyubov
transformations, is especially informative to study the process of filling the
Fermi sphere up because at this point the density of quark ensemble develops a
continuous dependence on the Fermi momentum. It allows us to reveal an
additional structure in the solution of the gap equation for dynamical quark
mass
just in the proper interval of parameters characteristic for phase transition
and
to trace its evolution. The result is that quark ensemble might be found in two
aggregate states, gas and liquid, and the chiral condensate is partially
restored
in a liquid phase. In order to make these conclusions easily perceptible we deal
with the simplest version of the NJL model (with one flavor and one of the
standard parameter sets) and, actually, are not targeted to adjust the
result obtained with well-known nuclear liquid-gas phase transition. Besides,
it seems our approach might be treated as a sort of microscopic ground of the
conventional bag model and those states filled up with quarks are conceivable
as a natural 'construction material' for baryons.

Now as an input to start with we remind the key moments of approach developed.
The corresponding model Hamiltonian includes the interaction term taken in the
form of a product of two colored currents located in the spatial points
${\bf x}$ and ${\bf y}$ which are connected by a formfactor and its density
reads as
\begin{equation}
\label{1}
{\cal H}=-\bar q(i{\bf \gamma}{\bf \nabla}+im)q-\bar q t^a\gamma_\mu q
\int d{\bf y}\bar q' t^b\gamma_\nu q' \langle A^{a}_\mu A'^{b}_\nu\rangle,
\end{equation}
where $q=q({\bf x})$, $\bar q=\bar q({\bf x})$, $q'=q({\bf y})$,
$\bar q'=\bar q({\bf y})$ are the quark and anti-quark operators,
\begin{eqnarray}
\label{2}
&&\hspace{-0.5cm}
q_{\alpha i}({\bf x})=\int\frac{d {\bf p}}{(2\pi)^3} \frac{1}{(2|p_4|)^{1/2}}
\left[a({\bf p},s,c)u_{\alpha i}({\bf p},s,c) e^{i{\bf p}{\bf x}}
+\right.\nonumber\\[-.2cm]
\\ [-.25cm]
&&~~~~\left.+b^+({\bf p},s,c)v_{\alpha i}({\bf p},s,c) e^{-i{\bf p}{\bf
x}}\right],
\nonumber
\end{eqnarray}
$p_4^2=-{\bf p}^2-m^2$, $i$--is the colour index, $\alpha$ is the spinor index
in the coordinate space, $a^+$, $a$ and $b^+$, $b$ are the creation and
annihilation operators of quarks and anti-quarks, $a~|0\rangle=0$,
$b~|0\rangle=0$,
$|0\rangle$ is the vacuum state of free Hamiltonian and $m$ is a current quark
mass. The
summation over indices $s$ and $c$ is meant everywhere, the index $s$ describes
two spin
polarizations of quark and the index $c$ plays the similar role for a colour. As
usual $t^a=\lambda^a/2$ are the generators of $SU(N_c)$ colour gauge group and
$m$ is
the current quark mass. The Hamiltonian density is considered in the Euclidean
space
and $\gamma_\mu$ denote the Hermitian Dirac matrices, $\mu,\nu=1,2,3,4$.
$\langle A^{a}_\mu A'^{b}_\nu\rangle$ stands for the formfactor of the following
form
\begin{equation}
\label{cor}
\langle A^{a}_\mu A'^{b}_\nu\rangle=\delta^{ab}
 \frac{2~\widetilde G}{N_c^2-1}\left[I({\bf x}-{\bf y})
\delta_{\mu\nu}-J_{\mu\nu}({\bf x}-{\bf y})\right],
\end{equation}
where the second term is spanned by the relative distance vector and the gluon
field primed denotes that in the spatial point ${\bf y}$. The effective
Hamiltonian density (\ref{1}) results from averaging the ensemble of quarks
influenced by intensive stochastic gluon field $A^a_\mu$, see Ref. \cite{MZ}.
For the sake of simplicity we neglect the contribution of the second
term in (\ref{cor}) in what follows. The ground state of the system is
searched as the Bogolyubov trial function composed by the quark-anti-quark
pairs with opposite momenta and with vacuum quantum numbers, i.e.
\begin{eqnarray}
\label{4}
&&\hspace{-0.65cm}|\sigma\rangle={\cal{T}}~|0\rangle~,~~~\nonumber\\[-.2cm]
\\ [-.25cm]
&&\hspace{-0.65cm}
{\cal{T}}=\Pi_{ p,s}\exp\{\varphi[a^+({\bf p},s)b^+(-{\bf p},s)+
a({\bf p},s)b(-{\bf p},s)]\}.\nonumber
\end{eqnarray}
In this formula and below, in order to simplify the notations, we refer to only
one complex index which means both the spin and colour polarizations.
The parameter $\varphi({\bf p})$ which describes the pairing strength is
determined by the minimum of mean energy
\begin{equation}
\label{5}
E=\langle\sigma|H|\sigma\rangle~.
\end{equation}
By introducing the 'dressing transformation' we define the creation and
annihilation operators of quasi-particles as
$A={\cal{T}}~a~{\cal{T}}^{-1}$, $B^+={\cal{T}}~b^+{\cal{T}}^{-1}$ and
for fermions ${\cal{T}}^{-1}={\cal{T}}^\dagger$.
Then the quark field operators are presented as
\begin{eqnarray}
\label{6}
&&q({\bf x})=\int\frac{d {\bf p}}{(2\pi)^3} \frac{1}{(2|p_4|)^{1/2}}~
\left[~A({\bf p},s)~U({\bf p},s)~e^{i{\bf p}{\bf x}}+\right.\nonumber\\
&&~~~~~~~\left.+B^+({\bf p},s)~V({\bf p},s)~ e^{-i{\bf p}{\bf
x}}\right]~,\nonumber\\
&&\bar q({\bf x})=\int\frac{d {\bf p}}{(2\pi)^3} \frac{1}{(2|p_4|)^{1/2}}~
\left[~A^+({\bf p},s)~\overline{U}({\bf p},s)~e^{-i{\bf p}{\bf
x}}+\right.\nonumber\\
&&~~~~~~~\left.+B({\bf p},s)~\overline{V}({\bf p},s)~ e^{i{\bf p}{\bf
x}}\right]~,
\nonumber
\end{eqnarray}
moreover, the transformed spinors $U$ and $V$ are given by the following forms
\begin{eqnarray}
\label{7}
&&U({\bf p},s)=\cos(\varphi)~u({\bf p},s)-
\sin(\varphi)~v(-{\bf p},s)~,\nonumber\\[-.2cm]
\\ [-.25cm]
&&V({\bf p},s)=\sin(\varphi)~u(-{\bf p},s)+
\cos(\varphi)~v({\bf p},s)~.\nonumber
\end{eqnarray}
where $\overline{U}({\bf p},s)=U^+({\bf p},s)~\gamma_4$,
$\overline{V}({\bf p},s)=V^+({\bf p},s)~\gamma_4$ are the Dirac conjugated
spinors.

In the paper Ref. \cite{ff} the process of filling in the Fermi sphere with the
quasi-particles of quarks was studied by constructing the state of the Sletter
determinant type
\begin{equation}
\label{8}
|N\rangle=\prod_{|{\mbox{\scriptsize{\bf P}}}|<P_F;S}~A^+({\bf
P};S)~|\sigma\rangle~,
\end{equation}
which possesses the minimal mean energy over the state $|N\rangle$. The
polarization indices run over all permissible values here and the quark
momenta are bounded by the limiting Fermi momentum $P_F$. The momenta and
polarizations of states forming the quasi-particle gas are marked by the
capital letters similar to above formula. In all other cases the small
letters are used.

As known the ensemble state at finite temperature $T$ is described by the
equilibrium statistical operator $\rho$. Here we use the Bogolyubov-Hartree-Fock
approximation in which the corresponding statistical operator is presented
by the following form
\begin{equation}
\label{dm}
\rho=\frac{e^{-\beta ~\hat H_{{\mbox{\scriptsize{app}}}}}}{Z_0}~,
~~Z_0=\mbox{Tr}~\{e^{-\beta ~\hat H_{{\mbox{\scriptsize{app}}}}}\}~~,
\end{equation}
where some approximating effective Hamiltonian
$H_{{\mbox{\scriptsize{app}}}}$ is
quadratic in the creation and annihilation operators of quark and anti-quark
quasi-particles $A^+$, $A$, $B^+$, $B$ and is defined in the corresponding Fock
space with the vacuum state $|\sigma\rangle$ and $\beta=T^{-1}$. We don't need
to know
the exact form of this operator henceforth because all the quantities of our
interest in the Bogolyubov-Hartree-Fock approximation are expressed by the
corresponding
averages (a density matrix)
$$n(P)=\mbox{Tr} \{\rho~ A^+({\bf P};S) A({\bf P};S)\}~,$$
$$\bar n(Q)=\mbox{Tr} \{\rho~ B^+({\bf Q};T) B({\bf Q};T)\}~,$$
which are found by solving the following variational problem. One needs to
determine the statistical operator $\rho$ in such a form in order
to have at the fixed mean charge
\begin{equation}
\label{ntot}
\bar Q_4=\mbox{Tr} \{\rho ~Q_4\}=
V ~2 N_c \int  \frac{d {\bf p}}{(2\pi)^3}~ [n(p)-\bar n(p)]~,
\end{equation}
where ($Q_4=-\int ~d{\bf x}~ \bar q i \gamma_4 q$)
$$
Q_4=\int\frac{d{\bf p}}{(2\pi)^3}\frac{-
ip_4}{|p_4|}\left[A^+(p)A(p)+B(p)B^+(p)\right],$$
for the diagonal component (of our interest here) and the fixed mean entropy
\begin{widetext}
\begin{eqnarray}
\label{stot}
&&\bar S=-\mbox{Tr} \{\rho \ln \rho\}=\nonumber\\[-.2cm]
\\ [-.25cm]
&&=-V 2 N_c \int\!\!\!  \frac{d {\bf p}}{(2\pi)^3}
\left[n(p)\ln n(p)+(1-n(p))\ln (1-n(p))
+\bar n(p)\ln \bar n(p)+(1-\bar n(p))\ln (1-\bar n(p))\right],\nonumber
\end{eqnarray}
\end{widetext}
($S=-\ln \rho$), the minimal value of mean energy of quark ensemble
$$E=\mbox{Tr} \{\rho~H\}~.$$
The definition of mean charge (\ref{ntot}) is given here up to the unessential
(infinite) constant coming from permuting the operators $B B^+$ in the
charge operator $Q_4$. It may not be out of place to remind that the mean
charge should be treated in some statistical sense because it characterizes
quark ensemble density and has no colour indices.

The contribution of free part of Hamiltonian
\begin{widetext}
\begin{eqnarray}
\label{12}
&&\hspace{-0.5cm}
H_0=-\int d {\bf x}~ \bar q({\bf x})~(i {\bf \gamma}
{\bf \nabla}+im)~q({\bf x})=\nonumber\\
&&\hspace{-0.5cm}=\int  \frac{d {\bf p}}{(2\pi)^3}|p_4|
\left[\cos \theta~ A^+({\bf p};s)A({\bf p};s)+\sin \theta~
A^+(-{\bf p};s)B^+({\bf p};s)+\right.\nonumber\\
&&\left.+\sin \theta ~B(-{\bf p};s)A({\bf p};s)-
\cos \theta~ B({\bf p};s)B^+({\bf p};s)\right],\nonumber
\end{eqnarray}
\end{widetext}
into the mean energy is calculated to be as
\begin{eqnarray}
\label{13}
&&\hspace{-1.cm}\mbox{Tr}\{\rho~{\cal H}_0\}=\int\frac{d {\bf p}}{(2\pi)^3}~
|p_4|~
(1- \cos\theta)+\nonumber\\[-.2cm]
\\ [-.25cm]
&&~~~~+\int\frac{d {\bf p}}{(2\pi)^3}~
|p_4| \cos\theta ~[n(p)+\bar n(p)]~,\nonumber
\end{eqnarray}
where ${\cal H}_0=H_0/V2 N_c$ is the specific energy and $\theta=2\varphi$.
It is obvious that a natural regularization has been done in the first term
of Eq. (\ref{13}) by subtracting the free Hamiltonian contribution $H_0$
(without pairing quarks and anti-quarks). In our particular case it is
quite unaffected to have the ensemble energy equal zero at the pairing angle
equal zero. It explains just an appearance of unit in the term
containing $\cos\theta$.

The interaction part of Hamiltonian, $\bar qt^a\gamma_\mu q\bar q't^a\gamma_\nu
q'$, provides four nontrivial contributions.
The term $ \mbox{Tr}\{\rho BB^+B'B'^+ \}$ generates the following components:
$\overline{V}_{\alpha i}({\bf p},s)t^a_{ij}\gamma^\mu_{\alpha \beta}
V_{\beta j}({\bf Q},T) \overline{V}_{\gamma k}({\bf Q},T)t^b_{kl}
\gamma^\mu_{\gamma \delta}V_{\delta l}({\bf p},s)$
(adding the similar term but with the changes $Q,T\to Q',T'$
which generates another primed quark current);
$-2\overline{V}({\bf Q},T)~t^a\gamma^\mu V({\bf Q}',T')
\overline{V}({\bf Q}',T')t^b\gamma^\mu V({\bf Q},T)$. Here
(as in the following expressions) we omitted all colour and spinor
indices which are completely identical to those of previous matrix element.
The term $ \mbox{Tr}\{\rho BAA'^+B'^+\}$ generates
the following nontrivial contributions:
$\overline{V}({\bf p},s)t^a\gamma^\mu U({\bf q},t)
\overline{U}({\bf q},t)t^b\gamma^\mu V({\bf p},s)$\\
$-\overline{V}({\bf p},s)t^a\gamma^\mu U({\bf P},S)
\overline{U}({\bf P},S)t^b\gamma^\mu V({\bf p},s)$\\
$-\overline{V}({\bf Q},T)t^a\gamma^\mu U({\bf q},t)
\overline{U}({\bf q},t)t^b\gamma^\mu V({\bf Q},T)$\\
$+\overline{V}({\bf Q},T)t^a \gamma^\mu U({\bf P},S)
\overline{U}({\bf P},S)t^b\gamma^\mu V({\bf Q},T)$.
Averaging $\mbox{Tr}\{\rho AA^+A'A'^+ \}$ gives the following terms:
$\overline{U}({\bf P},S)t^a\gamma^\mu U({\bf p},s)
\overline{U}({\bf p},s)t^b\gamma^\mu U({\bf P},S)$
(adding the similar term but with the changes $P,S\to P',S'$);
$-2\overline{U}({\bf P},S)t^a\gamma^\mu U({\bf P}',S')
\overline{U}({\bf P}',S')t^b\gamma^\mu V({\bf P},S)$.
And eventually the nontrivial contribution comes from averaging
$ \mbox{Tr}\{\rho A^+B^+B'A' \}$ and it has the form
$\overline{V}({\bf Q},T)t^a\gamma^\mu U({\bf P},S)
\overline{U}({\bf P},S)t^b\gamma^\mu V({\bf Q},T)$.
All other diagonal matrix elements generated by the following terms
$ \mbox{Tr}\{\rho AA^+B'B'^+\}$, $ \mbox{Tr}\{\rho BB'^+A'^+A' \}$,
do not contribute at all (their contributions equal to zero).
Similar to the calculation of matrix elements performed at zero temperature in
Ref. \cite{ff} we should carry out the integration over the Fermi sphere with
the corresponding distribution functions in the quark and anti-quark momenta
$\int^{P_F} \frac{d {\bf p}}{(2\pi)^3}$ $\to$ $\int
\frac{d {\bf p}}{(2\pi)^3}[n(p)+\bar n(p)]$
in our case of finite temperature. All necessary formulae for polarization
matrices which contain the traces of corresponding
spinors could be found in Refs. \cite{MZ} and \cite{ff}.
Bearing in mind this fact here we present immediately the result for
mean energy density per one quark degree of freedom as
$$w=\frac{{\cal E}}{2 N_c}~,~~~{\cal E}=E/V~,$$
where $E$ is a total energy of ensemble,
\begin{widetext}
\begin{eqnarray}
\label{14}
w&=&\int\frac{d {\bf p}}{(2\pi)^3} |p_4| \cos\theta [n(p)+\bar n(p)]+
2 G\int \frac{d {\bf p}}{(2\pi)^3}
\sin \left(\theta-\theta_m\right)[n(p)+\bar n(p)]
\int \frac{d {\bf q}}{(2\pi)^3} \sin\left(\theta'-\theta'_m\right)I-
\nonumber\\[-.2cm]
\\ [-.25cm]
&-&G\int \frac{d {\bf p}}{(2\pi)^3}~
\sin \left(\theta-\theta_m\right)[n(p)+\bar n(p)]
\int \frac{d {\bf q}}{(2\pi)^3}~
\sin\left(\theta'-\theta'_m\right)~[n(q)+\bar n(q)]~I+\nonumber\\
&+&\int \frac{d {\bf p}}{(2\pi)^3}~|p_4|(1-\cos\theta)
-G\int \frac{d {\bf p}}{(2\pi)^3}~\sin\left(\theta-\theta_m\right)
\int \frac{d {\bf q}}{(2\pi)^3}~\sin\left(\theta'-\theta'_m\right)~ I~,\nonumber
\end{eqnarray}
\end{widetext}
(up to the constant unessential for our consideration here).
In this formula the following denotes are used
$p=|{\bf p}|$, $q=|{\bf q}|$,  $\theta'=\theta(q)$, $I= I({\bf p}+{\bf q})$ and
the angle $\theta_m(p)$ is determined by the condition as follows
$$\sin \theta_m=\frac{m}{|p_4|}~.$$
It is interesting to notice that the existence of such an angle stipulates
the discontinuity of mean energy functional mentioned above and found out in
\cite{MZ}.
It is quite practical to single out the color factor in the four-fermion
coupling constant as $G=\frac{2 \widetilde G}{N_c}$.
Now let us make the following transformations while
integrating in the interaction terms
\begin{eqnarray}
&&\hspace{-0.5cm}
2 \int d {\bf p}f(p)\int d {\bf q} -\int d {\bf p} f(p)\int d {\bf q}f(q)
-\int d {\bf p}\int d {\bf q}=\nonumber\\
&&\hspace{-0.5cm}
=\int d {\bf p} f(p)\int d {\bf q}(1-f(q))-
\int d {\bf p}(1-f(p))\int d {\bf q},\nonumber
\end{eqnarray}
and changing the variables ${\bf p}$ $\leftrightarrow$
${\bf q}$ in the last term we obtain
\begin{eqnarray}
&&\int d {\bf p}~f(p)\int d {\bf q}~(1-f(q))-
\int d {\bf p}\int d {\bf q}~(1-f(q))=\nonumber\\
&&
-\int d {\bf p}~(1-f(p))\int d {\bf q}~(1-f(q))~.\nonumber
\end{eqnarray}
Finally we find out for the mean partial energy it looks like
\begin{widetext}
\begin{eqnarray}
\label{15}
&w&=\int\frac{d {\bf p}}{(2\pi)^3}~ |p_4| +
\int\frac{d {\bf p}}{(2\pi)^3}~|p_4|~\cos\theta~[n(p)+\bar n(p)-1]-\nonumber\\[-
.2cm]
\\ [-.25cm]
&-&G\int \frac{d {\bf p}}{(2\pi)^3}~\sin \left(\theta-\theta_m\right)~[n(p)+\bar
n(p)-1]
\int \frac{d {\bf q}}{(2\pi)^3}~\sin\left(\theta'-\theta'_m\right)~
[n(q)+\bar n(q)-1]~I~.\nonumber
\end{eqnarray}
\end{widetext}

We are interested in minimizing the following functional
\begin{equation}
\label{fun}
\Omega=E-\mu~\bar Q_4 -T~\bar S~,
\end{equation}
where $\mu$ and $T$ are the Lagrange factors for the chemical potential and
temperature respectively. The approximating Hamiltonian which we discussed above
$\hat H_{{\mbox{\scriptsize{app}}}}$, is constructed just by using the
information on $E-\mu~\bar Q_4$ of presented functional (see, also below).
For the specific contribution per one quark degree of freedom
$$f=\frac{F}{2N_c}~,~~~F=\Omega/V~,$$
we receive
\begin{widetext}
\begin{eqnarray}
\label{17}
f&=&\int \frac{d{\bf p}}{(2\pi)^3}~
\left[|p_4|\cos\theta~(n+\bar n -1)-\mu~(n-\bar n)\right]
+\int \frac{d{\bf p}}{(2\pi)^3}~|p_4|-\nonumber\\
&-&G\int \frac{d{\bf p}}{(2\pi)^3}~
\sin \left(\theta-\theta_m\right)~(n+\bar n-1)\int\frac{d{\bf q}}{(2\pi)^3}
~\sin\left(\theta'-\theta'_m\right)~(n'+\bar n'-1)~I+\\
&+&T\int \frac{d{\bf p}}{(2\pi)^3}~
\left[n~\ln n+(1-n)~\ln(1-n)+\bar n~\ln \bar n+(1-\bar n)~\ln(1-\bar
n)\right]~.\nonumber
\end{eqnarray}
\end{widetext}
Here the primed variables correspond to the momentum $q$. The optimal values of
parameters are determined by solving the following equation system
($df/d\theta=0$, $df/d n=0$, $df/d \bar n=0$)
\begin{eqnarray}
\label{18}
&&|p_4|~\sin\theta-M\cos \left(\theta-\theta_m\right)=0~,\nonumber\\[-.2cm]
\\ [-.25cm]
&&|p_4|~\cos\theta-\mu+M~\sin \left(\theta-\theta_m\right)-
T~\ln \left(\frac{1}{n}-1\right)=0~,\nonumber\\
&&|p_4|~\cos\theta+\mu+M~\sin \left(\theta-\theta_m\right)-
T~\ln \left(\frac{1}{\bar n}-1\right)=0~,\nonumber
\end{eqnarray}
where we denoted the induced quark mass as
\begin{equation}
\label{19}
\hspace{-0.5cm}
M({\bf p})=-2G\int\!\!\! \frac{d{\bf q}}{(2\pi)^3}
(n'+\bar n'-1)\sin \left(\theta'-\theta'_m\right)I({\bf p}+{\bf q}).
\end{equation}
\begin{figure}
\includegraphics[width=0.3\textwidth]{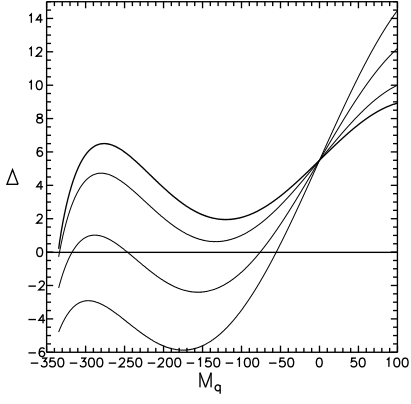}
\caption{The residual $\Delta$ for equation (\ref{19}) as a function of
dynamical quark mass $M_q$ (MeV) at zero value of temperature and the
following values of chemical potential $\mu$ (MeV) --- $335$
(the lowest curve), $340$, $350$, $360$ (the top curve).}
\label{f1}
\end{figure}

Turning to the presentation of obtained results in the form customary for mean
field approximation we introduce a dynamical quark mass $M_q$ parameterized as
\begin{equation}
\sin \left(\theta-\theta_m\right)=\frac{M_q}{|P_4|}~,~~
|P_4|=({\bf p}^2+M_q({\bf p}))^{1/2}~,
\end{equation}
and ascertain the interrelation between induced and dynamical quark masses.
From the first equation of system (\ref{18}) we fix the pairing angle
$$\sin \theta=\frac{p~M}{|p_4||P_4|}~,$$
and making use the identity
\begin{equation}
\label{iden}
(|p_4|^2-M ~m)^2+M^2 p^2=[p^2+(M-m)^2]~|p_4|^2~,
\end{equation}
find out that
$$\cos \theta=\pm\frac{|p_4|^2-m~M}{|p_4||P_4|}~.$$
For the clarity we choose the upper sign 'plus'. Then, as an analysis of the NJL
model teaches, the branch of equation solution for negative dynamical quark mass
is
the most stable one. Let us remember here we are dealing with the Euclidean
metrics
(though it is not a principal point) and a quark mass appears in the
corresponding
expressions as an imaginary quantity. Now substituting the calculated
expressions
for the pairing angle into the trigonometrical factor expression (obtained in
Ref. \cite{MZ})
$$\sin \left(\theta-\theta_m\right)=
\sin \theta~\frac{p}{|p_4|}-\cos \theta~\frac{m}{|p_4|}~.$$
and performing some algebraic transformations of both parts of equation we come
to the determination
\begin{equation}
\label{mass}
M_q({\bf p})=M({\bf p})-m~.
\end{equation}
And, in particular, the equation for dynamical quark mass (\ref{19}) is getting
the form characteristic for the mean field approximation
\begin{equation}
\label{23}
M=-2G~\int \frac{d{\bf q}}{(2\pi)^3}
~(n'+\bar n'-1)~\frac{M'_q}{|P'_4|}~I({\bf p}+{\bf q}).
\end{equation}

The second and third equations of ~system (\ref{18}) allow us to find for
the equilibrium densities of quarks and anti-quarks as
\begin{equation}
\label{newden}
n=\frac{1}{e^{\beta~(|P_4|-\mu)}+1}~,~~\bar n=\frac{1}{e^{\beta~(|P_4|+\mu)}+1}~,
\end{equation}
and, hence, the thermodynamical properties of our system as well, in particular,
the pressure of quark ensemble
$$P=-\frac{d E}{d V}~.$$
\begin{figure}
\includegraphics[width=0.3\textwidth]{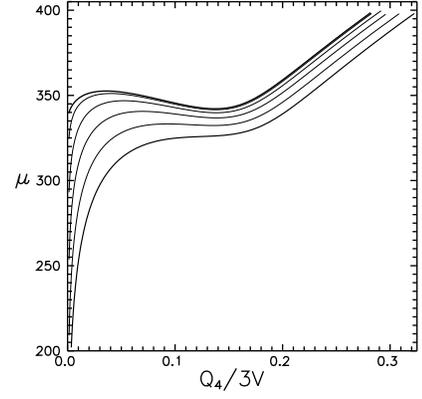}
\caption{The chemical potential $\mu$ (MeV) as a function of charge
density ${\cal Q}_4=\frac{Q_4}{3V}$ (in the units of charge/fm$^3$. The factor 3
relates the densities of quark and baryon matter. The top curve corresponds
to the situation of zero temperature. The curves following
down correspond to the temperature values $T=10$ MeV, ... , $T=50$ MeV with
spacing $T=10$ MeV.
}
\label{f2}
\end{figure}

By definition we should calculate this derivative at constant mean entropy,
$d\bar S/dV=0$. This condition allows us, for example, to calculate the
derivative $d\mu/dV$. However, this way is not reliable because then the mean
charge $\bar Q_4$ might change, and it is more practical to introduce two
independent chemical potentials --- for quarks $\mu$ and for anti-quarks
$\bar\mu$ (following Eq. (\ref{newden}) with an opposite sign). In fact,
it is the only possibility to obey both conditions simultaneously.
It leads to the following definitions of corresponding densities
$$n=\frac{1}{e^{\beta~(|P_4|-\mu)}+1}~,~~\bar
n=\frac{1}{e^{\beta~(|P_4|+\bar\mu)}+1}~.$$
In fact, this kind of description makes it possible
to treat even some non-equilibrium states of quark ensemble
(but with losing a covariance similar to the situation which takes place
in electrodynamics while one deals with electron-positron gas).
But here we are interested in the particular case of $\bar\mu=\mu$.
The corresponding derivative of specific energy
$\frac{d w}{d V}$ might be presented as
\begin{widetext}
$$
\frac{d w}{d V}=\int \frac{d{\bf p}}{(2\pi)^3}
\left(\frac{d n}{d\mu}\frac{d\mu}{dV}+\frac{d \bar n}{d\bar\mu}
\frac{d\bar\mu}{dV}\right)
\left[|p_4|\cos\theta-2G \sin \left(\theta-\theta_m\right)
\int\frac{d{\bf q}}{(2\pi)^3}
\sin\left(\theta'-\theta'_m\right)(n'+\bar n'-1)I\right].\nonumber
$$
\end{widetext}
Now expressing the trigonometric factors via dynamical quark mass
and exploiting Eq. (\ref{19}) we have for the ensemble pressure
\begin{equation}
\label{press}
\hspace{-0.15cm}P=-\frac{E}{V}-V2N_c\int \frac{d{\bf p}}{(2\pi)^3}
\left(\frac{d n}{d\mu}\frac{d\mu}{dV}+\frac{d \bar n}{d\bar\mu}
\frac{d\bar\mu}{dV}\right)
|P_4|.
\end{equation}

\begin{figure}
\includegraphics[width=0.3\textwidth]{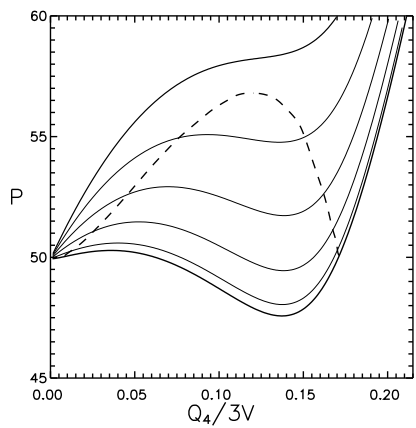}
\caption{The ensemble pressure $P$ (MeV/fm$^3$) as a function of charge density
${\cal Q}_4$ at temperatures $T=0$ MeV, ... , $T=50$ MeV with spacing $T=10$ MeV.
The lowest curve corresponds to zero temperature. The dashed curve shows the
boundary of phase transition liquid--gas, see the text.
}
\label{f2a}
\end{figure}
The requirement of mean charge conservation
\begin{equation}
\label{q4}
\hspace{-0.2cm}\frac{d \bar Q_4}{d V}=
\!\frac{\bar Q_4}{V}+V 2N_c\!\int \!\frac{d{\bf p}}{(2\pi)^3}
\left(\frac{d n}{d\mu}\frac{d\mu}{dV}-\frac{d \bar n}{d\bar\mu}
\frac{d\bar\mu}{dV}\right)=0,
\end{equation}
provides us with an equation which interrelates the derivatives $d\mu/dV$,
$d\bar\mu/dV$. Apparently, here the regularized expression for mean
charge of quarks and anti-quarks is meant (\ref{ntot}). Acting in
a similar way with the requirement of mean entropy conservation,
$d \bar S/dV=0$, we receive another equation as
\begin{eqnarray}
\label{entr}
&&\int \frac{d{\bf p}}{(2\pi)^3}\frac{d n}{d\mu}\ln \frac{n}{1-n}
\frac{d\mu}{d V} -\int \frac{d{\bf p}}{(2\pi)^3}
\frac{d \bar n}{d\bar\mu}\ln \frac{\bar n}{1-\bar n}
\frac{d\bar\mu}{d V}
=\nonumber\\[-.2cm]
\\ [-.25cm]
&&=\frac{\bar S}{2N_c~V^2}~.\nonumber
\end{eqnarray}
Substituting here $T~\ln\frac{n}{1-n}=\mu-|P_4|$ and
$T~\ln\frac{\bar n}{1-\bar n}=-\bar\mu-|P_4|$ after simple calculations
(keeping in mind that $\bar\mu=\mu$) we have with taking into
account (\ref{q4}) that
$$\int \frac{d{\bf p}}{(2\pi)^3}
\left(\frac{d n}{d\mu}\frac{d\mu}{dV}+\frac{d \bar n}{d\bar\mu}
\frac{d\bar\mu}{dV}\right)|P_4|=-\frac{\bar S T}{2N_c V^2}-\frac{\bar
Q_4\mu}{2N_c V^2}~.
$$
Finally it leads for the pressure to the following expression
\begin{equation}
\label{p}
P=-\frac{E}{V}+\frac{\bar S~T }{V}+\frac{\bar Q_4~\mu}{V}~.
\end{equation}
(of course, the thermodynamical potential is $\Omega=-P~V$).
At small temperatures the anti-quark contribution is negligible,
and thermodynamical description can be grounded on utilizing one
chemical potential $\mu$ only. If the anti-quark contribution
is getting intrinsic the thermodynamical picture becomes complicated
due to the presence of chemical potential $\bar\mu$ with the condition
$\bar\mu=\mu$ imposed. In particular, at zero temperature the anti-quark
contribution is absent and we might receive
$$P= -{\cal E}+\mu~\rho_q~,$$
where $\mu=\sqrt{P_F^2+M^2_q(P_F)}$, $P_F$ is the Fermi momentum
and $\rho_q=N/V$ is the quark ensemble density.
\begin{figure}
\includegraphics[width=0.3\textwidth]{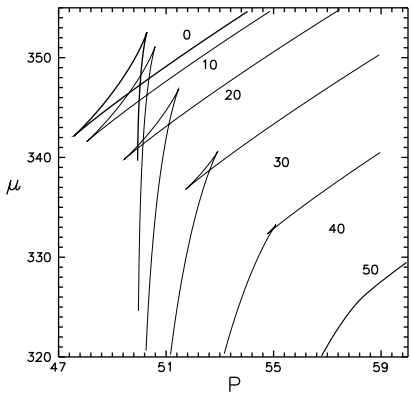}
\caption{The fragments of isoterms Fig. \ref{f2}, \ref{f2a}, see text.
Chemical potential $\mu$ (MeV) as a function of pressure $P$ MeV/fm$^3$.
The top curve corresponds to the zero isotherm and following down with
spacing 10 MeV till the isotherm 50 MeV (the lowest curve).}
\label{f3}
\end{figure}

For clarity, we consider mainly the NJL model \cite{Bub} in this paper,
i.e. the correlation function behaves as the $\delta$-function in coordinate
space. It is well known fact in order to have an intelligent result in this
model
one needs to use a regularization cutting the momentum integration in
Eq. (\ref{17}). We adjust the standard set of parameters \cite{5}
here with $|{\bf p}|<\Lambda$, $\Lambda=631$ MeV, $m=5.5$ MeV
and $G\Lambda^2/(2\pi^2)=1.3$. This set of parameters at $n=0$,
$\bar n=0$, $T=0$ gives for the dynamical quark mass
$M_q=335$ MeV. In particular, it may be shown the following
representation of ensemble energy is valid at the extremals
of functional (\ref{17})
\begin{eqnarray}
\label{mean}
&&E=E_{vac}+2 N_cV\int^\Lambda\frac{d{\bf p}}{(2\pi)^3}~|P_4|~(n+\bar
n)~,\nonumber\\
[-.2cm]
\\ [-.25cm]
&&E_{vac}=2 N_c V\!\!\! \int^\Lambda \frac{d{\bf p}}{(2\pi)^3}
(|p_4|- |P_4|)+2 N_c V \frac{M^2}{4 G},\nonumber
\end{eqnarray}
It is easy to understand this expression with the vacuum contribution
subtracted looks like the energy of a gas of relativistic particles
and anti-particles with the mass $M_q$, and
coincides identically with that calculated in the mean field approximation.

Let us summarize the results of this exercise. So, we determine the density
of quark $n$ and anti-quark $\bar n$ quasi-particles at given parameters
$\mu$ and $T$ from the second and third equations of system (\ref{18}).
From the first equation we receive the angle of quark and anti-quark
pairing $\theta$ as a function of dynamical quark mass $M_q$ which is
handled as a parameter. Then at small temperatures, below $50$ MeV,
and chemical potentials of dynamical quark mass order, $\mu\sim M_q$,
there are several branches of solutions of the gap equation. Fig. \ref{f1}
displays the difference of right and left sides of Eq. (\ref{19}) which is
denoted by $\Delta$ at zero temperature and several values of chemical
potential $\mu$ (MeV) = $335$ (the lowest curve),
$340$, $350$, $360$ (the top curve) as a function
of parameter $M_q$. The zeros of function $\Delta(M_q)$ correspond to the
equilibrium values of dynamical quark mass.

The evolution of chemical potential as a function of charge density
${\cal Q}_4=\frac{Q_4}{3V}$ (in the units of charge/$fm^3$) with
the temperature increasing is depicted in Fig. \ref{f2}
(factor 3 relates the quark and baryon matter densities). The top curve
corresponds to the zero temperature. The other curves following down have
been calculated for the temperatures $T=10$ MeV, ... , $T=50$ MeV with
spacing $T=10$ MeV. As it was found in Ref. \cite{ff} the chemical
potential at zero temperature is increasing first with the charge
density increasing, reaches its maximal value, then decreases and at the
densities of order of normal nuclear matter
density\footnote{At the Fermi momenta of dynamical quark mass order.},
$\rho_q\sim 0.16/fm^3$, becomes almost equal its vacuum value. Such a
behaviour of chemical potential results from the fast decrease of dynamical
quark mass with the Fermi momentum increasing. It is clear from this Fig.
the charge density is still multivalued function of
chemical potential at the temperature slightly below $50$ MeV.
The Fig. \ref{f2a} shows the ensemble pressure $P$ (MeV/fm$^3$)
as the function of charge density ${\cal Q}_4$ at several
temperatures. The lowest curve corresponds to the zero temperature.
The other curve following up correspond to the temperatures
$T=10$ MeV, ... , $T=50$ MeV (the top curve) with spacing
$T=10$ MeV. It is interesting to remember now that in Ref. \cite{ff}
the vacuum pressure estimate for the NJL model was received as $40$---$50$
MeV/fm$^3$ which is entirely compatible with the
results of conventional bag model. Besides, some hints at instability presence
(rooted in the anomalous behavior of pressure $dP/dn<0$) in an interval
of the Fermi momenta has been found out.
\begin{figure}
\includegraphics[width=0.3\textwidth]{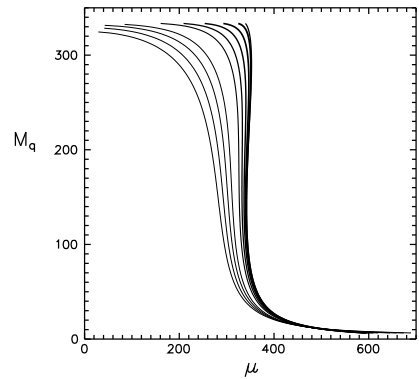}
\caption{The dynamical quark mass $|M_q|$ (MeV) as a function of chemical
potential $\mu$ (MeV) at the temperatures $T=0$ MeV, ... , $T=100$ MeV
with spacing $T=10$ MeV. The most right curve corresponds to zero temperature.
}
\label{f4}
\end{figure}

Fig. \ref{f3} shows the fragments of isotherms of Fig. \ref{f2}, \ref{f2a}
but in the different coordinates (chemical potential --- ensemble pressure).
The top curve is calculated at the zero temperature, the
other isotherms following down correspond to the temperatures increasing with
spacing 10 MeV. The lowest curve is calculated at the temperature 50 MeV.
This Fig. obviously demonstrate a presence of the states on isotherm which
are thermodynamically equilibrated and have equal pressure and
chemical potential (see the characteristic Van der Waals triangle with the
crossing curves). The calculated equilibrium points are shown in
Fig. \ref{f2a} by the dashed curve. The intersection
points of dashed curve with an isotherm are fixing the boundary of gas ---
liquid
phase transition. The corresponding straight line $P=\mbox{const}$ which
obeys the Maxwell rule separates the non-equlibrium and unstable fragments
of isotherm and describes a mixed phase. The corresponding
critical temperature for the parameter we are using in this paper turns out
to be $T_{c}\sim 45.7$ MeV with the critical charge density as
$\bar Q_4\sim 0.12$ charge/fm$^3$. Usually the thermodynamic
description is grounded on the mean energy functional which is the homogeneous
function of particle number like $E=N~f(S/N,V/N)$ (without vacuum contribution).
It is clear such a description requires the corresponding subtractions
to be introduced, however, this operation does not change the final
results considerably. It was argued in Refs. \cite{ff} that the states filled
up with quarks and separated from the instability region look like
'natural construction material' to form the baryons
and to understand the existing fact of equilibrium between vacuum and octet
of stable (in strong interaction) baryons\footnote{The chiral quark condensate
for the filled up state discussed develops the quantity about
(100 MeV)$^3$ (at $T=0$), see \cite{ff}, that demonstrates the obvious
tendency of restoring a chiral symmetry.}.
\begin{figure}
\includegraphics[width=0.3\textwidth]{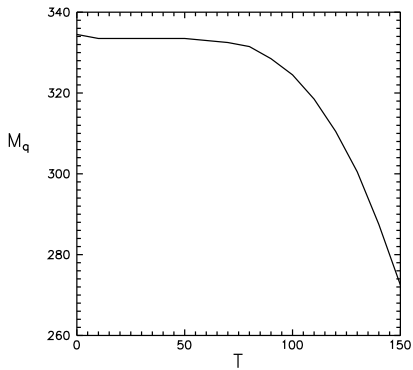}
\caption{ The dynamical quark mass $|M_q|$ (MeV) as a function of temperature
at small value of charge density ${\cal Q}_4$.}
\label{f4a}
\end{figure}

The dynamical quark mass $|M_q|$ (MeV) as a function of chemical potential
$\mu$ (MeV) is presented for the temperatures $T=0$ MeV, ... , $T=100$ MeV
with spacing $T=10$ MeV in Fig. \ref{f4}. The most right-hand curve
corresponds to the zero temperature. At small temperatures, below $50$ MeV,
the dynamical quark mass is the multivalued function of chemical potential.
The Fig. \ref{f4a} shows the dynamical quark mass as a function of temperature
at small values of charge density ${\cal Q}_4\sim 0$. Such a behaviour allows
us to conclude that the quasi-particle size is getting larger with temperature
increasing.
It becomes clear if we remember that the momentum corresponding the maximal
attraction between quark and anti-quark $p_\theta$ (according to
Ref. \cite{MZ}) is defined by $d\sin \theta/d p=0$. In particular, this
parameter in the NJL model equals to
\begin{equation}
\label{pt}
p_\theta=(|M_q|~m)^{1/2}~.
\end{equation}
but its inverse magnitude defines the characteristic (effective) size of quasi-
particle $r_\theta=p_\theta^{-1}$.

If one is going to define the quark chemical potential as an energy necessary
to add (to remove) one quasi-particle (as it was shown in \cite{ff} at zero
temperature), $\mu=dE/dN$, then in vacuum (i.e. at quark density $\rho_q$
going to zero) quark chemical potential magnitude coincides with the quark
dynamical mass. It results in the phase diagram displayed at this value
of chemical potential although, in principle, this value could be smaller
than dynamical quark mass as it has been considered in the pioneering
paper \cite{ayz}. If one takes, for example, chemical potential value equal
to zero it leads to the conventional picture but, obviously, such a
configuration
does not correspond to the real process of filling up the Fermi sphere with
quarks.

Apparently, our study of the quark ensemble thermodynamics produces quite
reasonable arguments to propound the hypothesis that the phase transition
of chiral symmetry (partial) restoration has already realized as the mixed
phase of physical vacuum and baryonic matter{\footnote{Indirect confirmation
of this hypothesis one could see, for example, in the existing degeneracy
of excited baryon states Ref. \cite{Gloz}.}}.
However, it is clear our quantative estimates should not be taken as ones to be
compared with, for example, the critical temperature of nuclear matter which has
been experimentally measured and is equal to 15 -- 20 Mev. Besides, the gas
component
(at $T=0$) has non-zero density (as $0.01$ of the normal nuclear density) but in
reality this branch should correspond to the physical vacuum, i.e. zero baryonic
density{\footnote{Similar uncertainty is present in the other predictions of
chiral
symmetry restoration scenarios, for example, it stretches from 2 to 6 units of
normal
nuclear density.}}. In principle, an idea of global equilibrium of gas and
liquid phases
makes it possible to formulate the adequate boundary conditions at describing
the
transitional layer arising between the vacuum and filled state and to calculate
the surface tension effects. We are planning to consider these aspects of phase
transition in a separate paper.

As a conclusion we would like to emphasize that in the present paper we
demonstrated how a phase transition of liquid--gas kind (with the resonable
values of parameters) emerges in the NJL-type models in which the quarks are
considered as the quasi-particles of the model Hamiltonian with four-fermion
interaction. The constructed quark ensemble displays some interesting features
for the nuclear ground state (for example, an existence of the state degenerate
with the vacuum one) but needs further study of its role in the context of
existing research \cite{KN} activity to explore the complicated
(or, may be, more realistic) versions of the NJL model and knowledge of the QCD
thermodynamics as gained in the lattice simulations.

The authors are deeply indebted K. A. Bugaev, R. N. Faustov, S. B. Gerasimov,
K. G. Klimenko, E. A. Kuraev, A.V. Leonidov, V. A. Petrov and A. M. Snigirev
for numerous fruitful discussions.


\bibliography{apssamp}

\end{document}